\documentclass[letter,11pt]{article}
\usepackage{amssymb}
\usepackage{amsmath}
\usepackage{amsfonts}
\usepackage{epsfig}
\usepackage{anysize}
\marginsize{3cm}{2cm}{2.5cm}{3cm}
\date{}

\begin{document}

\title{\LARGE \bf  Sums of variables at the onset of chaos, replenished}

\author{Alvaro D\'{i}az-Ruelas, Alberto Robledo\\ 
\footnotesize  Instituto de F\'{i}sica, Universidad Nacional Aut\' onoma de M\' exico y Centro de Ciencias de la Complejidad, \\ 
\footnotesize Apartado Postal 20-364, M\'exico 01000 DF, Mexico.\\
       }
			
\maketitle

\abstract{
As a counterpart to our previous study of the stationary distribution
formed by sums of positions at the Feigenbaum point via the period-doubling
cascade in the logistic map (\textit{Eur.\ Phys. J.} B \textbf{87} 32, (2014)), we
determine the family of related distributions for the accompanying cascade of
chaotic band-splitting points in the same system. By doing this we rationalize
how the interplay of regular and chaotic dynamics gives rise to either multiscale or gaussian limit distributions. As demonstrated before (\textit{J.\ Stat.\ Mech.}
\textbf{P01001} (2010)), sums of trajectory positions associated with the chaotic-band attractors of the logistic map lead only to a gaussian limit distribution, but,
as we show here, the features of the stationary multiscale distribution at the
Feigenbaum point can be observed in the distributions obtained from finite
sums with sufficiently small number of terms. The multiscale features are acquired
from the repellor preimage structure that dominates the dynamics toward the
chaotic attractors. When the number of chaotic bands increases this hierarchical
structure with multiscale and discrete scale-invariant properties develops.
Also, we suggest that the occurrence of truncated $q$-gaussian-shaped
distributions for specially prescribed sums are $t$-Student
distributions premonitory of the gaussian limit distribution.
} 

\maketitle

\section{Introduction}
\label{intro}

The stationary distributions of sums of consecutive positions of trajectories
generated by the logistic map have been found to follow a basic renormalization
group (RG) structure: a nontrivial fixed-point multiscale distribution at the period-doubling onset of chaos (Feigenbaum point) and a gaussian trivial fixed-point distribution for all chaotic attractors \cite{robledo1,robledo2}. At the Feigenbaum
point the limit distribution of sums of positions, generated by an ensemble of initial
conditions uniformly distributed in the full phase space, possesses an infinite-level
hierarchical structure that originates from the properties of the repellor set and its
preimages \cite{robledo3}. This ladder organization was elucidated \cite{robledo3}
through consideration of the family of periodic attractors, conveniently, the super-stable
attractors called supercycles \cite{schuster1}, along the period-doubling cascade. Here we complement
this study by considering instead the cascade of chaotic band-splitting attractors,
or Misiurewicz ($M_n, n=0,1,2,...$) points \cite{robledo4}. This gives us the opportunity
of analyzing the transformation of the developing multiscale distributions into the
gaussian limit distribution. The larger the number $2^n$ of bands in the attractor
the finer the approximation attained to the multiscale structure of the non-trivial fixed-point
limit distribution before it undergoes a crossover to the gaussian distribution.

\begin{figure}
\centering
\includegraphics[width=0.8\columnwidth]{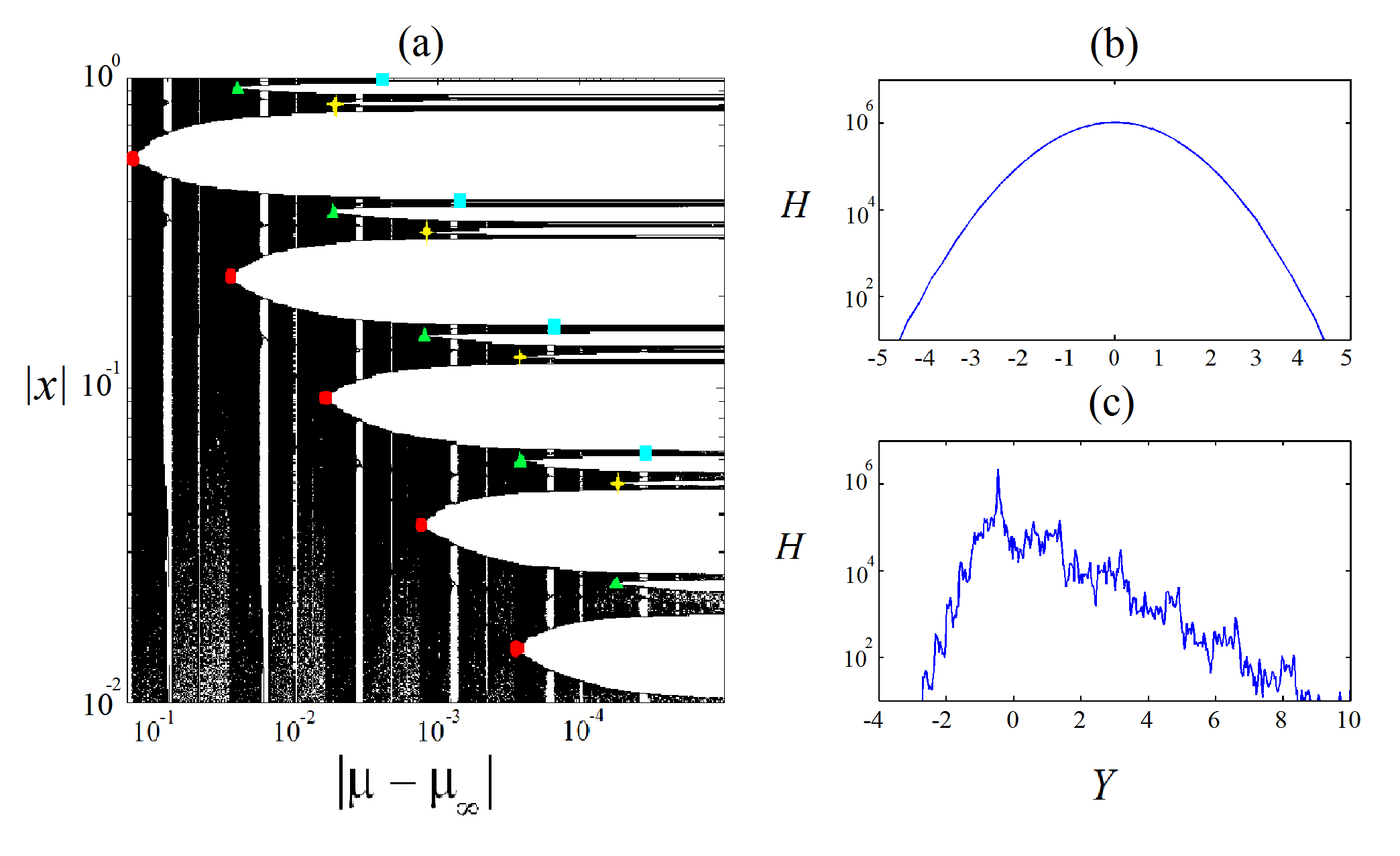} 
\caption{\footnotesize (Colour on-line.) (a) Attractor bands (in black) and gaps between them (white horizontal regions) in logarithmic scales, 
$-\log(\mu-\mu_{\infty })$ and $\log(|x|)$, in the horizontal and vertical axes, respectively. The band-splitting points $M_n$ (circle, triangle, diamond and square symbols) follow straight lines indicative of power-law scaling. The vertical white strips are periodic attractor windows. (b) Histogram $H$ of sums of trajectories when $\mu=2$ approaches the shape of the gaussian limit distribution. (c) Histogram of sums of trajectories at $\mu =$ $\mu _{\infty}$ approximates the shape of the multiscale limit distribution.}
\label{fig:1}       
\end{figure}

We consider the logistic map $f_{\mu }(x)=1-\mu x^{2}$, $-1\leq x\leq 1$,
$0\leq \mu \leq 2$, for which the control parameter value for its main period-doubling cascade accumulation point is $\mu =$ $\mu _{\infty}=1.401155189092 \ldots$
\cite{beck1}. When $\mu$ is shifted to values larger than $\mu _{\infty }$,
$\Delta \mu \equiv \mu -\mu _{\infty }>0$, the attractors are (mostly) chaotic and
consist of $2^{n}$ bands, $n=0,1,2,...$, where $2^{n}\sim \Delta \mu ^{-\kappa }$,
$\kappa =\ln 2/\ln \delta $, and $\delta =4.669201609102 \ldots$ is the universal
constant that measures both the rate of convergence of the values of $\mu =\mu _{n}$
to $\mu _{\infty }$ at period doubling or at band splitting points \cite{schuster1}. The
latter points are attractor merging crises, where multiple pieces of an attractor merge
together at the position of an unstable periodic orbit  \cite{grebogi1}. The  $M_{n}$
points can be determined by evaluation of the trajectories with initial condition
$x_{0}=0$ for different values of $\mu $,\ as these orbits follow the edges of the
chaotic bands until at $\mu =\mu _{n}$ the unstable orbit of period $2^{n}$ reaches
the merging crises \cite{grebogi1}. In Fig. \ref{fig:1}a we show (with different symbols) several
$M_{n}$ points that appear aligned because of the logarithmic scales employed.
The family of lines of equal slope along which the $M_{n}$ points fall reveal the
power-law scaling associated with their locations. In Fig. \ref{fig:1}b we show a corroboration
for the gaussian limit distribution for sums generated at the one-band fully-chaotic
attractor at $\mu=2$,  whereas in Fig. \ref{fig:1}c we show the multiscale limit distribution
for the Feigenbaum point at $\mu _{\infty }$.

Here we present results for ensembles of sums of positions and their distributions
at the attractors where bands split. On the one hand, we find that distributions obtained from
sums with an adequately finite number of terms resemble the asymmetric exponential-tailed shape of the limit distribution at $\mu _{\infty }$, and display gradually more of
its multiscale features as the order $n$ of the band-splitting point $M_n$ increases.
On the other hand, as the number of summands $N$ increases the distribution for
each value of $n$ develops a symmetrical shape and eventually approaches the
gaussian limit distribution for $N \rightarrow \infty$, as anticipated for all chaotic attractors.
We discuss the occurrence of so-called $q$-gaussian distributions for special sums of
positions at, or around, the $M_n$ points \cite{robledo4}, \cite{tsallis1}, \cite{tirnakli1}
in terms of distributions for finite-size data sets drawn from gaussian variables such
as it is the case of the $t$-Student distribution. 


\section{Sums of positions and their distributions for band-splitting attractors} \label{sec:2}

We consider the sum of consecutive positions $x_{t}$ up to a final iteration time $N$ of
a trajectory with initial condition $x_{0}$ and control parameter value $\mu$, \textit{i.e.}
\begin{equation}
X(x_{0},N;\mu )\equiv \sum\limits_{t=0}^{N}x_{t}
\label{eq:Xdef}
\end{equation}
for $x_0 \in [-1,1]$, as well as the centered, rescaled, sum
\begin{equation}
Y \equiv \frac{X - \left\langle X \right\rangle}{\sigma},
\label{eq:Ydef}
\end{equation}
where $\left\langle X \right\rangle$ is the average value for the sums over  the ensemble of $\mathcal{M}$ initial conditions, and $\sigma$ is the standard deviation of this ensemble. The use of $Y$ instead of $X$	 facilitates the comparison between sums with different values of $\mu_n$. We denote by $H$ the histogram corresponding to the sum $Y$. $H$ can be normalized or not, in which case it will be indicated. 

\begin{figure}
\centering
\includegraphics[width=0.8\columnwidth]{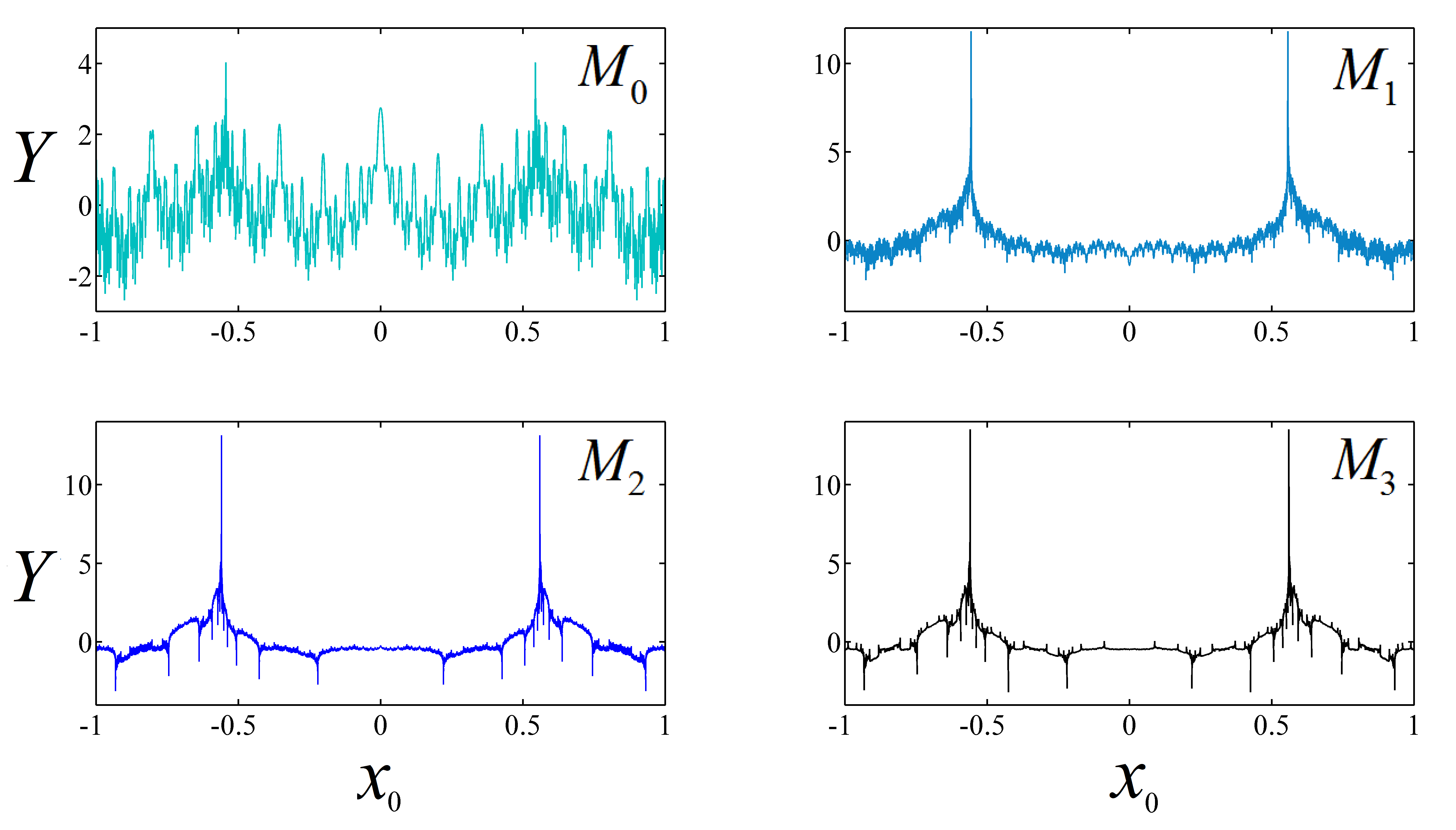}
\caption{\footnotesize Sums $Y(x_{0},N;\mu)$ as a function of the initial condition $x_0$.
These correspond, as indicated in each frame, to the first four Misiurewicz points, $M_0$,
$M_1$, $M_2$ and $M_3$, when the attractors are about to split into $2$, $4$, $8$ and
$16$ chaotic bands, respectively. The number of terms in the sums for $M_0$,
$M_1$, $M_2$ and $M_3$ are $N = 2^4, 2^5, 2^6, 2^7$, respectively. See text for description.}
\label{fig:2}       
\end{figure}

The four panels in Fig. \ref{fig:2} show the results for $X(x_{0},N;\mu )$ for all possible initial
conditions $-1\leq x_{0}\leq 1$ when the control parameter
takes the values for the first four Misiurewicz points, $M_n, n=0,1,2,3$. The
plots are all symmetrical with respect to $x_{0}=0$ and exhibit as a main feature
two matching peaks and a central valley. For $M_0$ the two peaks and the valley
are nearly concealed by other symmetrically-located peaks that form the fluctuating
background generated by the large chaotic band that forms the attractor. But as the number of
bands that form the attractor increases (and their widths decrease) for the
subsequent $M_1$, $M_2$ and $M_3$ points the twin large peaks and other, finer,
features become increasingly clear. This is the result of weaker intraband chaotic
motion fluctuations as the interband regular dynamics progressively dominates.

The panels in Fig. \ref{fig:2} can be compared with those in Fig. 2 in \cite{robledo3} where the
corresponding sums for the first supercycle attractors are shown. The same twin
peaks separated by a central valley and progressively finer motifs of alternating signs
appear in the sums of positions at the supercycle attractors \cite{robledo3}, only there
the absence of chaotic motion does not mask these details. In \cite{robledo3} it was demonstrated
that these features in the sums arise from the dynamics towards the attractors. Namely,
the two large peaks arise from the trajectories still close to the main repellor or its first
preimage. Whereas the finer peaks of alternating signs are due to trajectory positions near
subsequent repellors or their infinite families of preimages. See \cite{robledo3} for details.
Here we observe, as shown in Fig. \ref{fig:2} that for the Misiurewicz points the same developments
take place and that as the order $n$ of $M_n$ increases their effect on the sums becomes
increasingly clear.      

\begin{figure}
\centering
\includegraphics[width=0.8\columnwidth]{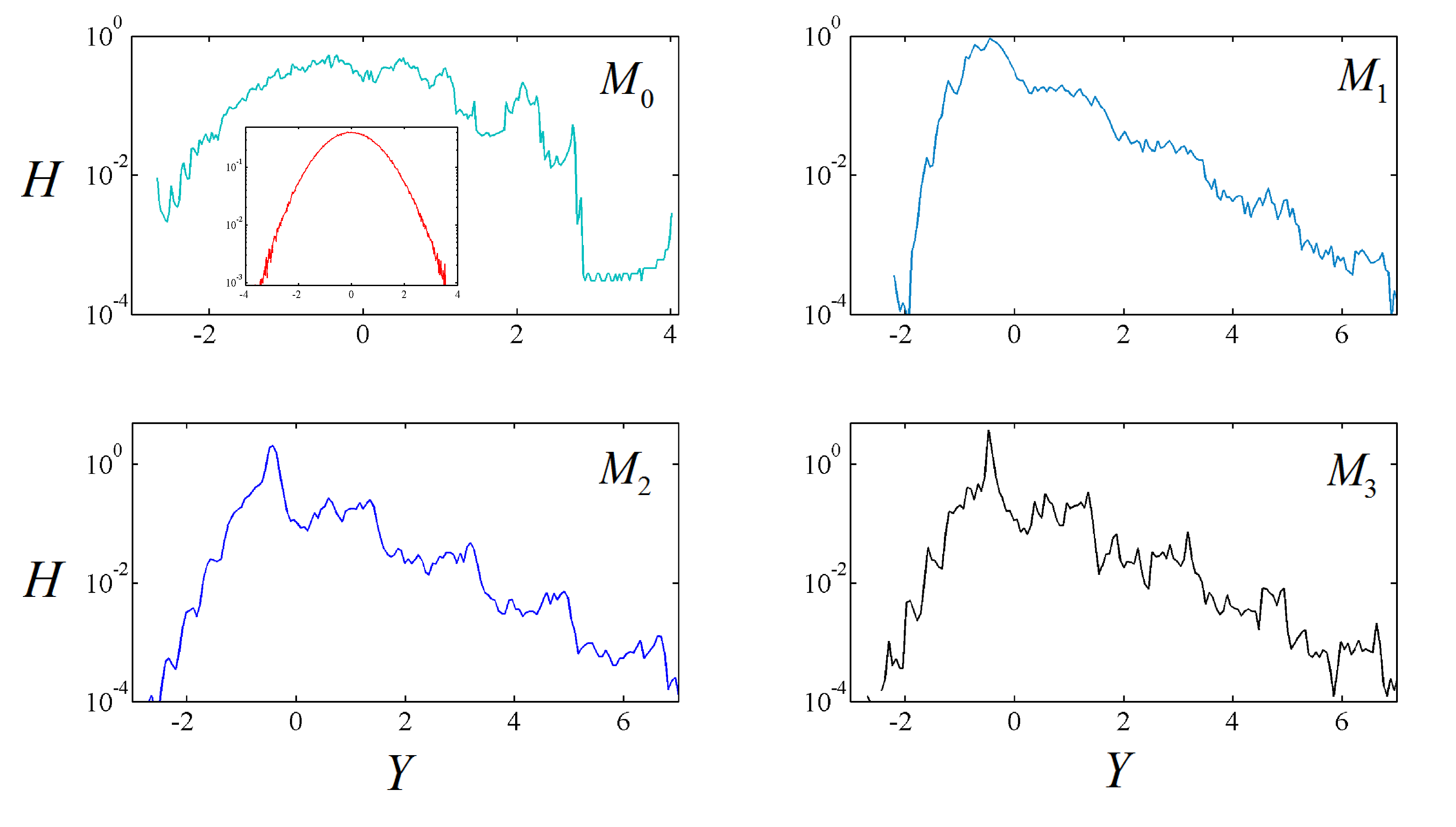}
\caption{\footnotesize Normalized histograms $H$, obtained from the sums in Fig. \ref{fig:2}. As indicated, these correspond to the
first four Misiurewicz points, $M_n, n=0,1,2,3$. In the inset for $M_0$ the sum consists of $O(2^{10})$ terms. See text for description}
\label{fig:3}       
\end{figure}

We look now at the distributions associated with the sets of sums $X(x_{0},N;\mu )$
shown previously. The four panels in Fig. \ref{fig:3} present in semi-logarithmic scales
the results for the (normalized) histograms that correspond to the sums
in the panels for the first four Misiurewicz points, $M_n, n=0,1,2,3$ in Fig. \ref{fig:2}. In the
case of $M_0$ one obtains an almost round shape with some uneven, pointy, features at the right hand side. 
In the inset we show the attainment of a gaussian form when the number of summands is increased.
The histogram for the next point $M_1$ already shows a basic tent shape formed by a sharp rise followed by a
gentler decline. The following case $M_2$ displays serrated features over the tent
shape that appear even more pronounced in the last case for $M_3$. Thus, as $n$
increases we observe the development of the asymmetrical double-exponential
global shape, with superimposed motifs of ever-decreasing finer detail, of the
stationary distribution for $\mu_{\infty }$ shown in Fig. \ref{fig:1}c.

\section{Truncated $q$-gaussian ($t$-Student) crossover distributions at band-splitting attractors}

Special sums of positions of ensembles of trajectories within the chaotic band attractors
of the logistic map generate distributions that are well reproduced by truncated $q$-gaussian
shapes \cite{tsallis1}, \cite{tirnakli1}. These sums are generated by first discarding an opening
set of terms $N_{s}$ such that the ensemble of trajectories initiated as a uniform distribution
within $[-1,1]$ is already located within the multiple-band attractor \cite{robledo4}. Secondly,
the sums are stopped at a total number of terms $N_{f}$ such that their distribution still displays
large non-exponential, but truncated, tails. That is, these kind of sums are of the form

\begin{equation}
X(x_{0},N_{s},N_{f};\mu)\equiv \sum\limits_{t=N_{s}}^{N_{s}+N_{f}}x_{t}.
\label{Eq:sums}
\end{equation}.

\begin{figure}[t!]
\centering
\includegraphics[width=0.8\columnwidth]{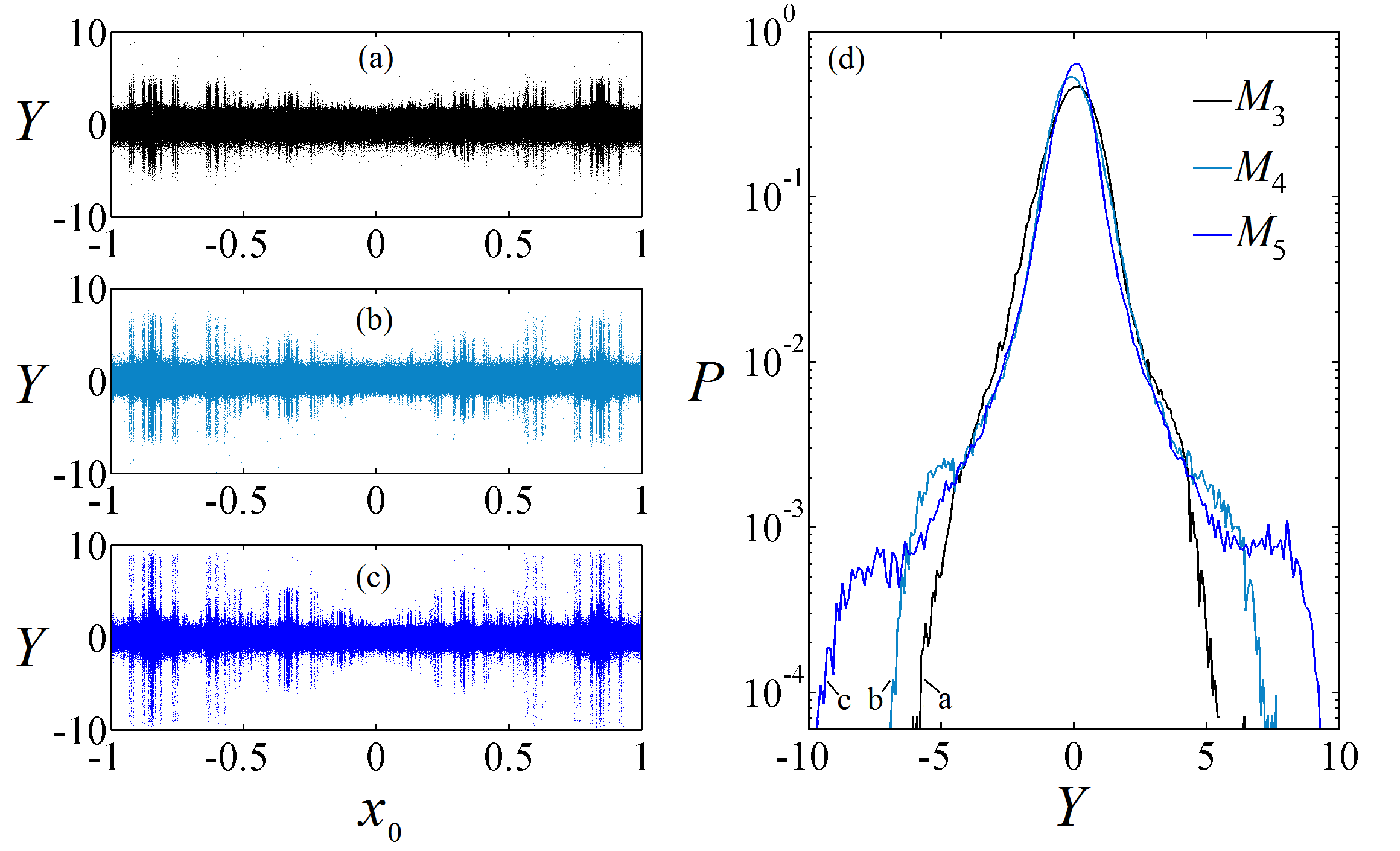} 
\caption{\footnotesize (Colour on-line.) Rescaled sums obtained from a uniform distribution of $10^6$
initial conditions across $[-1,1]$ at the band-splitting points $M_n, \ n= 3,4,5 $ with labels
(a),(b),(c), respectively, and with their corresponding distributions (normalized histograms) $P$ in (d). The values of
$N_s$ and $N_f$  used are, respectively $N_s = 2^6,2^7,2^8,$ and  $N_f = 2^{9}, 2^{10}, 2^{11}$.}
\label{fig:4}       
\end{figure}

In Fig. \ref{fig:4} where we show in panels (a), (b) and (c) the centered sums
$Y \equiv (X - \langle X \rangle)/\sigma$ for the Misiurewicz points $M_3, \ M_4$ and $M_5$, when
the attractors are about to split into $16$, $32$ and $64$ chaotic bands, respectively.
In panel (d) we show the distributions $P$ for these sums without rescaling of the
horizontal axis $Y$. This kind of distributions have been fitted by $q$-gaussians
\cite{tsallis1}, \cite{tirnakli1}.

The Student's distribution or $t$-distribution is defined through the random variable 
\begin{equation}
\vartheta = \frac{W\sqrt{n}}{\sqrt{Z}},
\label{eq:tDistRV}
\end{equation}
where $n$ is the sample size, the random variable $X$ follows a gaussian distribution and $Z$ is a random variable $Z = Z_1^2 + Z_2^2 + Z_3^2 + \ldots + Z_n^2 $ following a $\chi^2$-distribution. The sum in Eq. (\ref{Eq:sums}) when it is centered and rescaled as in Eq. (\ref{eq:Ydef}), can be viewed as a random variable similar to that  in Eq. (\ref{eq:tDistRV}) by including the crucial and delicate, but rather well-justified, assumption that the sums of positions follow a gaussian distribution. This is possible provided that the correlation function for the positions $x_t$ goes to zero as the number of iterations $N$ goes to infinity in a chaotic regime, as it is certainly fulfilled by the collection of Misiurewicz points we have used \cite{beck1}. Under this assumption, the quantity in the numerator $\tilde{X} \equiv X-\langle X \rangle$ will (asymptotically) follow a gaussian distribution, and, considering the fact that the standard deviation in Eq. (\ref{eq:Ydef}) is $\sigma = (1/\mathcal{M}\sum_i^\mathcal{M}(X_i-\langle X\rangle))^{1/2}$  we can write 
\begin{equation}
Y = \frac{\tilde{X}\sqrt{\mathcal{M}}}{\sqrt{\tilde{Z}}},
\end{equation}
where we have considered $\tilde{Z}$ to be $\tilde{Z}=\sum_i^\mathcal{M} \tilde{X_i}$.
The probability density function of the variable defined in Eq. (\ref{eq:tDistRV}) is
\begin{equation}
P(\vartheta) = \frac{\Gamma\left(\frac{n+1}{2}\right)}{(n\pi)^{1/2}\Gamma(\frac{n}{2})}\left( 1+\frac{\vartheta^2}{n} \right)^{-\frac{n+1}{2}}.
\label{eq:tDistpdf}
\end{equation}
So we can expect, and then corroborate numerically, that our variable $Y$ follows the same distribution, hence the sums of iterates $Y$ can be regarded to follow a $t$-distribution. It is well known that the $t$-distribution arises when estimating the mean of
finite data sets drawn from an infinite set that pertains to a gaussian distribution \cite{student2}. The larger the sample the closer the $t$-distribution resembles the gaussian distribution and, in the limit $n\rightarrow\infty$ this is no longer an approximation and we recover a Gaussian, as we can check straightforwardly from Eq. (\ref{eq:tDistpdf}).

It has been demonstrated \cite{tsallis2} that this distribution optimizes the Tsallis entropy
expression \cite{tsallis3}. The expression for the $t$-distribution coincides (after a suitable
change of variable) with that of the so-called $q$-gaussian distribution \cite{tsallis3}
\begin{equation}
P(y) = \left(\frac{\beta(q-1)}{\pi}\right)^{\frac{1}{2}}\frac{\Gamma \left( \frac{1}{q-1}\right)}{\Gamma \left( \frac{1}{q-1}-\frac{1}{2}\right)}(1+\beta(q-1)y^2)^{\frac{1}{1-q}}, \ \ 1<q<3.
\end{equation}

Therefore, the appearance of $q$-gaussian-like distributions for these families of sums
are predictable as the chosen finite sets of data cannot still approximate the gaussian
limit distribution. Interestingly, the $t$-Student (or $q$-gaussian) distribution maximizes
an entropy expression that satisfies only three of the Khinchin axioms (with the exclusion of composability) \cite{turner1} whereas
the gaussian distribution maximizes the canonical entropy expression that satisfies all four
Khinchin axioms.


\section{Summary and discussion}


We have probed the dynamical properties of chaotic attractors of the logistic map with
the intention of learning about the distributions formed by sums of ensembles of positions of trajectories
with uniformly-distributed initial conditions through the total phase space $[-1,1]$. For
suitability we chose members of the family of attractors at which bands split. Our results are
consistent with the existence of a single type of limit distribution (i.e. when the number
of summands $N \rightarrow \infty$) for all chaotic attractors, this being the gaussian
distribution. But we also observed, when the sums have a sufficiently small number of
terms, the characteristic features of the multiscale, asymmetric, exponential-tailed,
limit distribution that occurs for the special case of the accumulation point of the band-splitting attractors, the Feigenbaum point at $\mu _{\infty }$. The crossover that takes
place from distributions displaying the multiscale form to the gaussian shape was also
addressed. Sums obtained when all trajectories are placed within the chaotic bands
and are built-up from finite sets of positions that evolve randomly are comparable to sets
of variables extracted from a gaussian population as this is their limit distribution. Such
procedure generates the $t$-Student distribution, equivalently a $q$-gaussian distribution,
as previous studies detected.  

\section*{Acknowledgments}
AD-R and AR acknowledge support from DGAPA-UNAM-IN103814 and CONACyT-CB-2011-167978
(Mexican Agencies).


\begin{thebibliography}{99}

\bibitem{robledo1} Fuentes M.A. and Robledo A.  
                   \textit{J.\ Stat.\ Mech.} \textbf{P01001}
                (2010).

\bibitem{robledo2} Fuentes M.A. and Robledo A.              
		\textit{J.\ Phys.\ Conf.\ Ser.} \textbf{201}, 012002
		(2010).               
\bibitem{robledo3} Fuentes M.A. and Robledo A. 
                   \textit{Eur.\ Phys. J.} B \textbf{87}, 32 
                (2014).
\bibitem{schuster1} Schuster R.C.
                    \textit{Deterministic Chaos. An Introduction.} VCH Publishers, Weinheim
                    (1988).
                    

\bibitem{robledo4} Diaz-Ruelas A., Fuentes M.A. and Robledo A. 
                   \textit{Europhys.\ Lett.} \textbf{108}, 20008 
                (2014).
								
\bibitem{beck1} Beck C. and Schlogl F.  
                \textit{Thermodynamics of Chaotic Systems.} Cambridge University Press, Cambridge
                (1993).
                   
\bibitem{grebogi1} Grebogi C., Ott E. and Yorke J. A. 
                   \textit{Physica} D \textbf{7}, 181
                (1983).
\bibitem{tsallis1} Tirnakli U., Tsallis C. and Beck C.,
		 \textit{Phys. Rev.} E \textbf{75}, 040106(R) (2007); \textit{Phys. Rev.} E 79, 056209 (2009).            

\bibitem{tirnakli1} Afsar O. and Tirnakli U.,
		 \textit{Europhys. Lett.} \textbf{101}, 20003 (2013); \textit{Physica} D \textbf{272}, 18 (2014).

\bibitem{student1} DeGroot, M.H., \textit{Probability and Statistics} Addison-Wesley Publishing Company, Reading, Massachusetts (1986) 
                
\bibitem{student2} Feller, W. \textit{An Introduction to Probability Theory and its Applications Vol. II} 2nd ed., John Wiley \& Sons, New York (1971)

\bibitem{tsallis2} de Souza, A.M.C., Tsallis, C.,
		\textit{Physica} A \textbf{236}, 52 (1997).

\bibitem{tsallis3} Tsallis, C., \textit{Introduction to Nonextensive Statistical Mechanics:
		 Approaching a Complex World}, Springer (2009).
		 


\bibitem{turner1} Hanel R., Thurner S., Gell-Mann M.,
                \textit{Proc. Natl. Acad. Sci.} USA \textbf{108}, 6390–6394 (2011).

\end{thebibliography}
\end{document}